# Improving Human Decisions by Adjusting the Alerting Thresholds for Computer Alerting Tools According to User and Task Characteristics

Marwa Gadala[1], Lorenzo Strigini[1], Peter Ayton[2]
[1] Centre for Software Reliability
City, University of London
London, United Kingdom

[2] Centre for Decision Research
University of Leeds
Leeds, LS2 9JT
United Kingdom
{Marwa.Gadala.1, L.Strigini}@city.ac.uk; P.Ayton@leeds.ac.uk

Technical report

Version 1.1, June 2021

**Abstract**

**Objective:** To investigate whether performance (number of correct decisions) of humans supported by a computer alerting tool can be improved by tailoring the tool's *alerting threshold* (sensitivity/specificity combination) according to user ability and task difficulty.

**Background:** Many researchers implicitly assume that for each tool there exists a single ideal threshold. But research shows the effects of alerting tools on decision errors to vary depending on variables such as user ability and task difficulty. These findings motivated our investigation.

**Method:** Forty-seven participants edited text passages, aided by a simulated spell-checker tool. We experimentally manipulated passage difficulty and tool alerting threshold, measured participants' editing and dictation ability, and counted participants' decision errors (false positives + false negatives). We investigated whether alerting threshold, user ability, task difficulty and their interactions affected error count.

**Results:** Which alerting threshold better helped a user depended on an interaction between user ability and passage difficulty. Some effects were large: for higher ability users, a more sensitive threshold reduced errors by 30%, on the easier passages. Participants were not significantly more likely to prefer the alerting threshold with which they performed better.

**Conclusion:** Adjusting alerting thresholds for individual users' ability and task difficulty could substantially improve effects of automated alerts on user decisions. This potential deserves further investigation. Improvement size and rules for adjustment will be application-specific.

**Application:** Computer alerting tools have critical roles in many domains. Designers should assess potential benefits of adjustable alerting thresholds for their specific CAT application. Guidance for choosing thresholds will be essential for realizing these benefits in practice.

**Précis:** *Computer alerting tools* (CATs) are useful and ubiquitous. A CAT's effects depend on its *alerting threshold* (sensitivity/specificity combination). This study suggests that adjusting CAT thresholds, according to the skills of the user and the difficulty of decisions required, has the potential to substantially improve user decisions.

**Improving Human Decisions by Adjusting the Alerting Thresholds for Computer Alerting Tools According to User and Task Characteristics**

*Computer alerting tools* (CATs) are a ubiquitous category of decision support systems with critical roles in disparate domains such as medical screening, collision alarms in vehicles, and airport luggage security screening. Every CAT operates as part of a larger system, with a typical setup as in Figure 1. In this setup, the CAT and human form a single socio-technical system (outer rectangle), and both the CAT and human have access to relevant information, but the final decision lies with the human.

**Figure 1** *Typical setup of human-CAT systems*

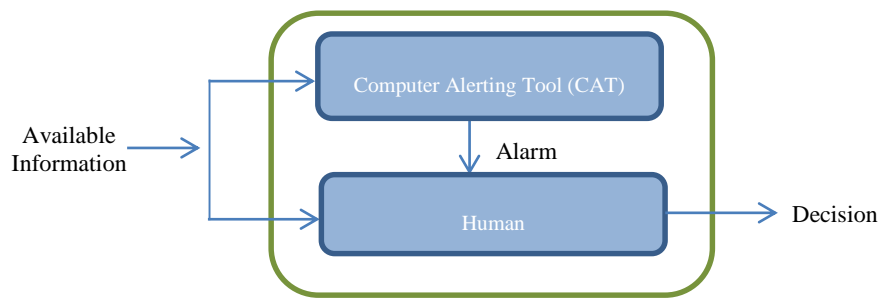

Despite their differences, all CATs detect given targets and alert the user to them via prompts (also called warnings, alarms). CAT prompts may be correct - either *true positive (TP)* or *true negative (TN)*; or incorrect - either *false positive (FP)* or *false negative (FN)*. Users weigh the CAT's output together with their own judgement to yield the overall decision. Similar to the CAT's output, the final output of the socio-technical system (human decision) is a: TP, TN, FP, or FN.

No CAT is perfect. The inevitable trade-off between the goals of high sensitivity to targets and low false-alarm rates is determined by the CAT setting. This can be visualized in terms of an alerting threshold: the CAT assesses the probability of something being a target, and issues a prompt if it exceeds the threshold. Many CAT designs implicitly assume a single ideal threshold exists for each CAT, but some have user-adjustable thresholds. Examples include patient monitoring devices for physiological parameters (Meyer & Sheridan, 2017; RGB, 2019) and Hologic's (2010) CAT for breast cancer detection allowing nine different setting combinations (sensitivity/false prompt rates for microcalcifications and masses). The following relevant bodies of literature are discussed below: choice of appropriate alerting thresholds; "automation bias" and how CATs can cause automation-mediated errors; adaptive/adaptable automation; and adjustable alerting thresholds.

**Choice of Alerting Threshold**

Which alerting threshold is best, given the inevitable trade-off between the frequencies of CAT FNs and FPs? Signal detection theory gives a criterion for choosing the optimal threshold as a function of the FN and FP rates and their costs. But the true usefulness of a CAT is determined by its actual effect on the user's final decision (Sorkin & Woods, 1985; Strigini, Povyakalo & Alberdi, 2003) because optimizing the alerting threshold in isolation, as if the CAT made the decision alone, can cause suboptimal human decision-making. For example, imagine a smoke alarm optimized to detect minimal amounts of smoke to reduce risk of deaths. The resulting high false alarm rate might induce users to ignore alarms, resulting in worse overall risk (i.e., more FN decisions by users) than if the detector were less sensitive (Sorkin et al., 1985). In the next section we address diverse ways that CATs can harm human decisions.

## CATs and Automation-Mediated Errors

A large body of literature documents how CATs may cause automation-mediated errors: making certain decision errors more likely and/or introducing decision errors that don't occur without CAT. We call such undesired effects "automation bias", warning, however, that the referent of this term varies between authors. Some assume automation bias has one possible cause and define it accordingly as "complacency", "overtrust" or "overreliance" (Goddard et al., 2011; Mosier et al., 1998) – which risks neglecting other important causes (Alberdi et al., 2009).

How can CATs cause errors? It is intuitive that CAT FNs can cause human FNs – as documented even in experts who, without CAT, would have detected the target (Alberdi et al., 2009). Improving the CAT's *sensitivity* (i.e. reducing CAT FNs) would mitigate this effect, but also usually increase CAT FPs (reduce *specificity*). Thus, one must also consider effects of CAT FPs. Appropriately, in some applications, like patient monitoring, there is growing recognition of CAT FPs as a major cause of dangerous errors (Bach, Berglund & Turk, 2018).

### Effects of CAT FPs

Controlled studies of CAT FP effects usually highlight reduced specificity of human decisions: CAT FPs increase human FPs. Fewer studies also consider sensitivity. Zheng et al. (2001) noted, in mammography reading, that increasing CAT FP rates made users miss more indications of possible cancer in non-prompted areas. If a CAT is calibrated for high sensitivity, the resulting high FP rate may produce worse decision accuracy than if the CAT were set to emit fewer alarms, because users may learn to ignore alarms (Sorkin et al., 1985); users may interpret absence of a prompt as indicating absence of a target (Alberdi et al., 2010); excessive/obvious FPs may distract users from targets, or make them discount CAT TPs (Alberdi et al., 2010); a target may occur in a location where FPs habitually occur (Hartswood et al., 1998); or the FPs may cause "subsequent search misses" (Biggs, 2017) and/or increased cognitive load (Alberdi et al., 2009; Taylor et al., 2008a; Lawrence et al., 2010). Furthermore, all CAT errors may reduce trust[1] in the CAT (Dzindolet et al., 2003; Lawrence et al., 2010). However, not all CAT FPs are harmful: prompting non-targets that resemble targets can aid users understand CAT capabilities and design rationale (Hartswood, Procter, Williams, Prescott & Dixon, 1996) and sometimes FPs can help train users' reactions to prompts (Parasuraman, Hancock & Olofinboba, 1997).

---

[1] We use "trust" in its general English meaning, but note that there is a large area of research concerning the conceptualization of this variable and factors that influence it (e.g., Hoff and Bashir, 2015; Moray et al., 2000).

## Effects of user and task characteristics

Research on automation bias shows that effects of CAT advice on user decisions depend on both user and task characteristics (Povyakalo et al., 2013). There is evidence that inexperienced users are especially prone to automation-mediated errors (Taylor et al., 2008a). Possible explanations include: (1) experienced users are more confident, rely less on CAT advice, and/or perceive the CAT as less trustworthy, especially after experiencing CAT errors (Dzindolet et al., 2003), and (2) experienced users can better recognize CAT errors, thus reducing their influence.

Nonetheless, automation bias can also affect experienced users (Dreiseitl & Binder, 2005), sometimes even more than inexperienced users (Mosier et al., 1998; Taylor et al., 2008b). Some studies also show higher-ability users (experience is not necessarily synonymous with ability) suffering from automation bias more often than lower-ability users (Galletta et al., 2005; Povyakalo et al., 2013) - especially for certain classes of errors (e.g., CAT FNs). Galletta et al. (2005) attributed this to laziness or "overreliance" in higher-ability users. We suggested alternative possible causes (Alberdi et al., 2009): with hard-to-detect targets, higher-ability users may hesitate, then use CAT FNs to (erroneously) adjudicate, but lower-ability users would miss such targets anyway, so CAT FNs here do not harm them. Povyakalo et al. (2013) corroborate this explanation, reporting greater benefit from CAT for lower ability users than for higher ability users, but mostly for easier targets; for difficult targets, CAT degraded the sensitivity only of the higher ability users. For another scenario in which user proficiency has unexpected drawbacks with alarms see Meyer & Bitan, 2002.

Since CATs' effectiveness varies with user ability/experience and task difficulty, we wondered whether adjusting alerting thresholds according to these two variables would reduce automation-mediated errors. Research about improving CATs by considering individual differences is already prominent in human factors and ergonomics, and Parasuraman & Manzey (2010) note that extending this research to address automation bias would be worthwhile. A related argument by Galletta et al (2003) is that "consonance" among users' cognitive skills, the task and its representation affects CAT effectiveness. Other researchers have questioned whether vehicle collision alerts should be adjusted according to driver capabilities/age (Parasuraman et al., 1997) or the sensitivity of medication management tools should be different for older users (Ho et al., 2005) or, more generally, whether CAT settings should be tailored to each user (Alberdi et al., 2010). Such questions deserve further attention.

## Adaptive/Adaptable Automation

A growing body of literature suggests that some drawbacks of automation, including complacency, lack of situational awareness, skill degradation, and mistrust can be alleviated by making automation adjustable during use, under user control ("adaptable automation") or automated control ("adaptive automation") (Kidwell et al., 2012). Several studies in this area focus on dynamically altering the tasks delegated to automation ("level of automation", LOA), rather than alerting thresholds (our present topic). Inagaki and Sheridan (2012) offer a probabilistic argument for this approach. Others compare the effects of adaptable and adaptive LOAs on system performance (Kidwell et al., 2012; Ruff et al., 2018).

**Adjustable Alerting Thresholds**

Several studies have evaluated CATs with ranges of alerting thresholds (Lehto et al., 2000; Zheng et al., 2001). Notably, Zheng et al. showed that lower CAT reliability significantly decreased user performance - to the extent that sometimes users performed better without CAT.

Adjustable thresholds in CATs promise several benefits. Adjustability can accommodate users' diverse needs. Users freed from preset settings will probably feel empowered, improving user satisfaction, as found by Lawrence et al. (2002). User-adjustable thresholds also allow organizations to diffuse responsibility, for certain failures, to the users - for better or for worse (Meyer & Sheridan, 2017). We study adjustable thresholds more generally inasmuch as they may be set, not just by individual users, but by vendors or user organizations, and we consider their potential for improving decisions.

Introducing adjustable thresholds prompts the question: how should an appropriate threshold be selected? Presently, some CATs are marketed with adjustable thresholds but without guidelines for choosing them. Moreover, relying on user preferences may not guarantee good choices. Botzer et al. (2010) noted that people generally adjusted the threshold in the right direction, but not magnitude. Lehto et al. (1998) found participants adjusted the threshold in the wrong direction approximately 30% of the time. Lawrence et al. (2002) showed that users' suboptimal selection of CAT settings produced less accurate decisions overall than those obtained using non user-adjustable CATs. Our study addresses this issue by considering factors that should be taken into account when choosing appropriate alerting thresholds.

## Current Study

Based on the prior research results outlined above, one may expect effects of user ability, task difficulty and alerting threshold on performance, which we define in terms of the frequency of user decision errors (FNs and FPs). A less obvious and more interesting issue is whether these factors interact and if so, how they interact.

Ideally, with adequate knowledge of how these factors affect users' performance, it would be possible to tune a CAT, before operation, to have the best threshold for its intended user and circumstances of use. This is a complex problem, as we now illustrate with an example.

Consider a collision alert CAT for drivers. To simplify this example, imagine that there are only two alerting thresholds: one which has a higher FP rate than the other in return for a lower FN rate. The CAT's designers may reason that because novice drivers often miss signs of danger, they would drive most safely with the highly sensitive CAT threshold. Instead, expert drivers who readily detect most dangers would drive most safely with the more specific CAT threshold, because frequent CAT FPs could cause fatigue and/or "cry wolf" effects (drivers ignoring TP alarms). This is one, plausible theory.

However, another plausible theory offers an entirely contradictory prescription. In this theory, designers might reason that CAT FPs, which experts effortlessly ignore, often cause novices to overreact dangerously – e.g. brake suddenly. On balance, then, safety is best served by giving novices a more specific threshold than experts. Both theories rely on effects that have been documented to occur; what matters for deciding the appropriate threshold to apply is which one has the dominant effect in these circumstances of use. In other words, which pairing of thresholds to driver ability is best depends on many factors such as the frequencies of hard-to-detect obstacles, of hard braking upon CAT FNs, etc. – unknown before experimentation (Gadala, 2017).

Moreover, the choice of appropriate alerting threshold not only depends on user ability, but is also likely affected by other factors such as task difficulty. Suppose that designers experimentally determine that for experts, in dry weather, the more specific CAT threshold should be used. In wet weather, with poorer visibility, experts may need CAT alerts for a larger subset of the potential dangers; while slippery roads make the novices' jittery responses to CAT FPs more dangerous. The respective thresholds that served novices and experts best for dry, summer weather may thus need to be swapped in wet, winter months.

To generalize from this example, choosing personalized alerting thresholds for a specific CAT is complex; it depends on a number of factors, and to decide, one needs empirical results from the testing of that specific CAT on samples of users, in representative situations. Armed with these results, designers could then extrapolate guidelines for appropriate alerting thresholds, e.g., a trucking company rules for setting thresholds for individual drivers.

In this study, we experimentally investigate whether adjusting a CAT's alerting threshold according to both user ability and task difficulty can improve decision accuracy. As in the example above, we had a detailed theory of how "easy" or "difficult" targets and CAT errors may affect users of varying abilities (Gadala, 2017); this predicted that the overall direction of the effects would depend on the relative frequencies of decisions of various degrees of difficulty for the specific sample of participants - unknown before the experiment. We thus designed this first study to (1) reveal just the *presence* of effects, irrespective of direction, because this would indicate the potential of this approach to improve CAT effectiveness, and suggest investing in detailed empirical measurements for each CAT; (2) investigate the complexity of the phenomenon: not just the effects of ability and difficulty alone, but how these factors *interact*.

Following our reasoning above, we hypothesize:

H1: User ability and alerting threshold will interact such that higher ability users will decide most accurately using one alerting threshold, and lower ability users when using another alerting threshold.

H2: The interaction in H1 will be moderated by a second order interaction with task difficulty: for each user ability group, the alerting threshold that yields better decision accuracy for difficult tasks may be different from the alerting threshold that yields better decision accuracy for easy tasks.

We note that "ability" and "difficulty" are generic terms. To make threshold adjusting practical in a specific CAT application, measures must be chosen that are feasible to obtain in that context. For instance, in various areas of clinical radiology it is feasible to assess individual radiologists' FN and FP rates in detecting targets, and the generic difficulty indicators for X-ray images.

## Method

As a target CAT application, we chose spell-checking. The benefit of low-cost experimentation coupled with a high degree of experimental freedom make it attractive for preliminary studies judging the usefulness of future studies on more critical CAT applications. Furthermore, spell-checking is an example of a broad category of visual search tasks on static images.

These may be very different in detail, but we see both theoretical reasons and some empirical corroboration for believing that the lessons learned on one CAT application will be beneficial for others. The theoretical reason is that all these tasks amount to detecting patterns and responding to them. The details of pattern detection – even the brain regions involved – may well be very different between these different tasks. But the required function of integrating strength of impression of the pattern being present with the frequency with which various situations (FNs and FPs especially) have been experienced and other information, towards satisfying a requirement for good performance, is common to all; and the mechanisms through which the mind performs this task in the various applications may well be similar. Hence, if experience with one CAT application helps us to hypothesize a set of mechanisms that affect user performance with that CAT, conjecturing similar mechanisms in the use of other CATs is a useful starting point, although their collective effect will need to be ascertained empirically for each specific CAT, as the balance between mechanisms operating in opposite directions will depend on situation-specific factors. Experience corroborates – to some extent – this argument that similar cognitive mechanisms with similar effects may operate in all these tasks. For example, Galletta et al. (2005) found behavior patterns in spell-checking matching those found by Alberdi et al. (2010) in mammography reading: in both studies, higher ability users were more likely to miss targets missed by the CAT, compared to when not using the CAT, while lower ability users were harmed less.

**Participants**

To achieve 80% power for the effect size observed by Galletta et al. (2005), we recruited forty-seven participants through email invitations and posters hung around our university. Each participant was paid one £8 Amazon voucher, and as an incentive for high performance, all participants producing correct text gained entry into a £50 prize draw. Participants gave informed consent through an online questionnaire requesting demographic and other information, such as their first language, years of spoken English, self-assessed spelling ability, and number of non-academic books read per year.

**The Task**

We used a simulated spell checker CAT and a within participant design. The CAT highlighted, in red, some words as misspelled, without suggesting corrections so as not to create any confounding factors. Participants were instructed: "correct as many SPELLING errors as you can before the time expires". We briefed participants that this advanced spell checker was programmed to detect non-words, and also misused words, especially homophones (e.g. "there" versus "their"); and that it was not 100% accurate so it might sometimes incorrectly highlight correctly spelled words, or fail to highlight mistakes.

**Procedure**

After participants registered online, they attended an in-person experiment session where they completed the *editing task*, the *dictation task* and several *questionnaires* (see Figure 2). The study was approved by the Research Ethics Committee at City, University of London which fully implements the Declaration of Helsinki principles.

**Figure 2** *Participant journey through the experiment*

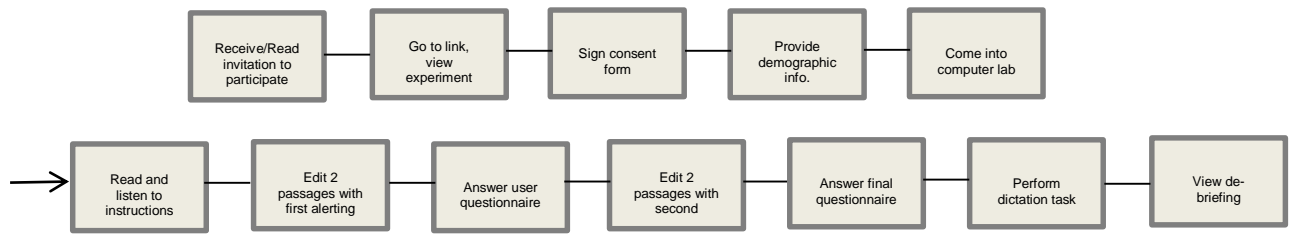

### The Editing Task

Each participant edited four passages: one "easy" and one "difficult" (defined *infra*) with a sensitive alerting threshold, and another "easy" and another "difficult" passage with a more specific alerting threshold. Order of texts and alerting thresholds were each independently randomized. The two alerting thresholds were presented as two different CATs, but editing instructions remained the same. Based on pilot observations, and to avoid floor and ceiling effects, we allocated participants 6 minutes to edit easy passages and 9 minutes to edit difficult passages.

### The Dictation Task

To assess spelling ability independently of the editing task, participants completed a self-paced dictation. They heard thirty words (repeated by the computer as many times as they wished), and then typed their spelling of each word. Words were from the list by Fischer et al., (1985, p. 424) which tests people's "linguistic sensitivity": "ability to apprehend the inherent regularities at various levels of linguistic representation". Fischer et al. identify three main levels of linguistic sensitivity, related to: (1) regular phonetic-to-orthographic mapping of words (such as in "retort"), (2) established orthographic conventions and abstract morphophonemic structures (such as the doubling of the consonant in "thinned"), and (3) patterns that rarely occur in English (such as in "bourgeois").

### The Questionnaires

Questionnaires interleaved in the editing task measured, on a 5-point Likert scale, participants' perception of each "CAT's" behavior and reliability, and trust in each "CAT". In a final questionnaire, participants also indicated whether they noticed any differences between the numbers of FPs and FNs produced by each "CAT", and with which "CAT" they preferred to work. The latter was used to investigate whether, given the choice, users could identify the alerting threshold that was better for them.

### Text Passages and Alerting Thresholds

For the editing task, we wrote four passages of, on average, 650 words about a range of subjects such as "How to Save a Wet Phone" or "Habits of Highly Effective People". We used Fischer et al.'s work to classify errors by difficulty, and ensured that "difficult" passages contained more errors of linguistic sensitivity 2 and 3. Misspellings were generated mostly by switching often-confused patterns rather than creating obvious keyboard errors (e.g., misspelling "disbelieve" as "disbeleive", not "fisbelieve").

Table 1 summarizes, by passage difficulty, the number of manipulated words of each level, the number misspelled, and the proportion marked by each CAT.

**Table 1**

*Distribution of Words Selected for Manipulating the CAT Threshold (Target Words) and CAT FN/FP errors*

| Passage Difficulty | Alerting Threshold | Target Word Type | Level 3 | | Level 2 | | Level 1 | | Homophones | |
|---|---|---|---|---|---|---|---|---|---|---|
| | | | Marked | Unmarked | Marked | Unmarked | Marked | Unmarked | Marked | Unmarked |
| | (Total Target Words: Misspelled, Correctly Spelled) | | (6: 3, 3) | | (6: 2, 4) | | (12: 1, 11) | | (14: 7, 7) | |
| | Sensitive Threshold | Misspelled | 3 | 0 | 2 | 0 | 1 | 0 | 7 | 0 |
| Easy passages | | Correctly spelled | 2 | 1 | 3 | 1 | 2 | 9 | 2 | 5 |
| | Specific Threshold | Misspelled | 2 | 1 | 2 | 0 | 1 | 0 | 2 | 5 |
| | | Correctly spelled | 1 | 2 | 1 | 3 | 0 | 11 | 1 | 6 |
| | (Total Target Words: Misspelled, Correctly Spelled) | | (12: 6, 6) | | (12: 4, 8) | | (None: 0, 0) | | (14: 7, 7) | |
| | Sensitive Threshold | Misspelled | 6 | 0 | 4 | 0 | 0 | 0 | 7 | 0 |
| Difficult passages | | Correctly spelled | 4 | 2 | 6 | 2 | 0 | 0 | 2 | 5 |
| | Specific Threshold | Misspelled | 4 | 2 | 4 | 0 | 0 | 0 | 2 | 5 |
| | | Correctly spelled | 2 | 4 | 2 | 6 | 0 | 0 | 1 | 6 |

*Note.* All counts are per passage. CAT FN errors generated with the sensitive threshold are a subset of those with the specific threshold and FP errors with the specific threshold are a subset of those with the sensitive threshold. The percentage of words misspelled in Levels 1-3 (10%; 30%; 50%) are, for ecological validity, approximated from the performance of participants in Fischer et al.'s study. The number of target words is the same between easy and difficult passages, but difficult passages contain more words of Levels 2 and 3, difficult for both the user (according to Fischer et al. (1985)) and the CAT. Hence, for ecological validity, the CAT's sensitivity was manipulated separately for each word level and for homophones (sensitive threshold: 100% sensitivity for Levels 1-3 and for homophones; specific threshold: 66.7% sensitivity for Level 3, 100% sensitivity for Level 2 and approximately 30% sensitivity for homophones). For passages of a given difficulty, the overall number of CAT errors was the same between the two thresholds, although, by definition, they were distributed differently between CAT FNs and FPs.

## Types of Decision Errors

For both the editing and dictation tasks we counted errors in the output produced by each participant. "Errors" in editing were tallied using two different criteria (see Table 2): the "Correct Detection" criterion disregards whether the user's correction is "appropriate" (defined in Table 2), "Correct Result" only considers the latter. For example, a user who correctly detects a misspelling, but fails to correctly rectify it, commits an error in result but not detection. Alternatively, a user who changes a correctly spelled and appropriate word commits an error in *detection*, even if the change results in an appropriate and correctly spelled word: a correct *result*.

**Table 2**

*Classification of Decision Outcomes*

|  | Definition of True Positive Decision | Definition of False Positive Decision |
|---|---|---|
| Classified by Correct Result | User changes a misspelled word to a correctly spelled and appropriate word | User changes a correctly spelled and appropriate word to a misspelled word |
| Classified by Correct Detection | As above<br><br>OR<br><br>User unsuccessfully attempts to correct a misspelled word<br><br>*e.g., "ellicit" to "elicit" rather than "illicit"* | As above<br><br>OR<br><br>User changes a correctly spelled and appropriate word to another correctly spelled and appropriate word<br><br>*e.g., "preventive" to "preventative" or "afflict" to "affect"* |

[a] "misspelled" is used to describe a word that does not exist in English or one that does not fit the context (e.g. *there* for *their*); hence, the complement of "misspelled" is "correctly spelled and appropriate".

This distinction between detection and result matters. For instance, in the context of mammography, Alberdi et al. (2008) found that 13.5% of recall decisions in their study were due to "features other than those indicating actual cancer", such as a radiologist correctly recalling a patient who has a mass in the right breast, but stating there is a problem in the left breast. Most mammography studies would classify these "Detection TPs" as TPs, but, in reality, a biopsy on the wrong breast would likely produce a "Result FN".

## Measures of Ability

Recall that to measure user ability we introduced the dictation task. Another proxy for ability is the total count of errors in the editing task over all conditions. Error counts in dictation and in editing likely reflect related but different aspects of ability – indeed previous studies indicate differences between the "knowledge deficit" hypothesis (users failing to correct a misspelling because they lack *knowledge* of the word's correct spelling) and the "processing deficit" hypothesis (users failing to correct a misspelling because they fail to *detect* the error) (Figueredo & Varnhagen, 2004). Another key difference between the two abilities is that editing ability includes ability to recognize CAT FPs.

We measured dictation ability by counting the words a participant misspelled in the dictation; our measure of editing ability counted the sum of errors (FPs + FNs by Correct Result: Table 2) remaining in all four passages after editing. Galletta et al. (2005) defined two ability groups, *high* and *low verbals,* using a median split. We followed their approach when graphing or discussing results.

## Results

### Participant Scores

Table 3 summarizes demographics of our participants.

**Table 3**

*Demographics of Study Participants*

| | |
|---|---|
| Total number of participants | 47 |
| Gender: Number of females | 37 (79%) |
| Occupation: Number of students | 36 (77%) |
| First Language: English | 26 (55%) |
| Self-Reported Spelling Ability | |
|   Poor | 1 (2%) |
|   Average | 6 (13%) |
|   Good | 28 (60%) |
|   Excellent | 12 (26%) |
| Mean Age (S.D.) | 30 (8) |
| Mean Years Speaking English (S.D.) | 22 (11) |

Dictation and editing scores both met tests for normality. Dictation scores ranged from 9 to 30 (out of 30), with mean 21.3, median 21, and *SD* 4.8. We analyzed the effect of spelling ability, using the Mann-Whitney U test for nominal variables, 1-Way ANOVA for ordinal variables, and Pearson's correlation for continuous variables. Factors significantly associated with participants' dictation score were: (1) first language (English vs Other) $U = 142$, $p = 0.005$, (2) number of non-academic books read per year $F(4,42) = 4.13$, $p = 0.007$, (3) self-assessed spelling ability $F(3,43) = 7.07$, $p = 0.001$ and (4) years of spoken English $r(47) = 0.510$, $p < 0.001$. These results reveal expected relationships and indicate no reasons for mistrusting the dictation scores.

As expected, we found high positive correlation between dictation and editing abilities $(r(47) = 0.765, p < 0.001)$, yet the two scores were not perfectly correlated. Thus, when presenting the remaining results, we separately analyze our hypotheses according to both ability criteria.

Table 4 summarizes the total decision errors of participants in the various conditions.

**Table 4**

*Mean (S.D.) Number of Remaining Errors by Condition*

| Alerting Threshold | Passage Difficulty | |
|---|---|---|
| | Easy | Difficult |
| Sensitive Threshold | 6.79 (3.13) | 11.6 (4.10) |
| | Min = 0; Max = 15 | Min = 3; Max = 21 |
| Specific Threshold | 7.47 (2.58) | 12.0 (4.50) |
| | Min = 2; Max = 12 | Min = 1; Max = 20 |

*Note.* Number of remaining errors defined according to Correct Result (cf Table 2)

Although, as discussed in the Methods section, we had reasons to expect differences between dictation and editing abilities, fine-grained analysis of the participants' decision errors (FP vs. FN) gave us further insights. High verbals, defined by editing ability, had, by definition, fewer total remaining errors than low verbals. On the other hand, high verbals, defined by dictation ability, corrected significantly more misspellings $(t(45) = 5.06, p < 0.001)$ than low verbals, but were no better at avoiding FPs $(t(45) = 0.247, p = 0.806)$. Thus, although better scores in both dictation and editing ability associated with correcting more misspelled words (blue, vertical stripes in Figure 3), higher scores in dictation were not associated with fewer FPs (red, horizontal stripes in Figure 3).

**Figure 3** *Comparison of error types in the editing task according to different definitions of user ability*

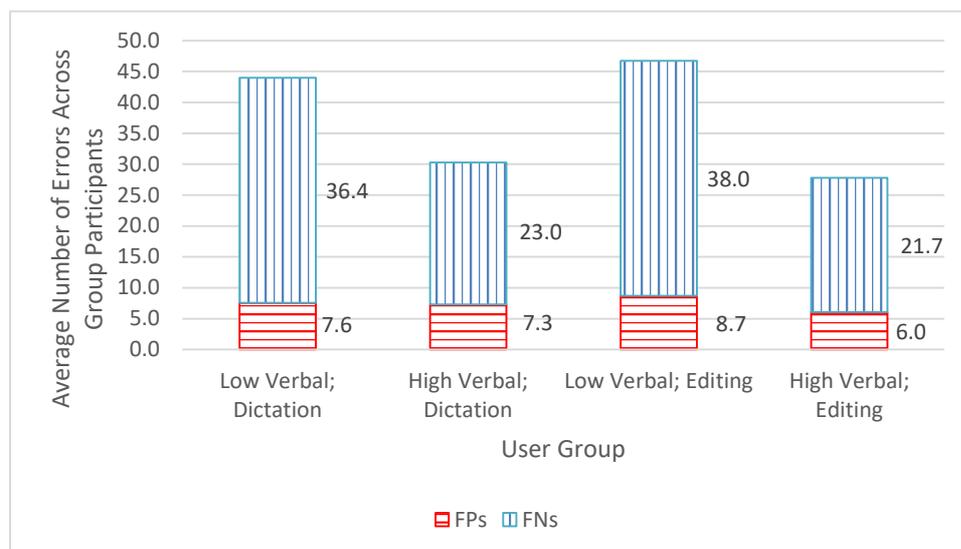

## Results addressing Hypothesis 1: Interaction between user ability and alerting threshold

We tested H1 via a repeated-measures ANOVA that compared *total remaining errors* in the passages between the two threshold settings (sensitive, specific threshold), and included ability as a co-variate (which avoids invidious criteria for distinguishing discrete levels of ability). The ANOVA was repeated eight times (Table 5), to study the interaction of interest at each difficulty level (easy vs difficult passages), and according to the two possible measures of ability (dictation vs editing) and criteria for correctness (Correct Result vs Correct Detection).

The ANOVA on the total remaining errors when editing easy passages, according to Correct Result, shows a significant interaction between alerting threshold and user Editing ability ($F(1, 45) = 4.46$, $p = .040*$, $\eta^2 = .09$; see Table 5). This supports Hypothesis 1: that performance depends on the interaction between user ability and alerting threshold. Nevertheless, the corresponding interaction using Dictation ability and all interactions for the difficult passages are not statistically significant. Although the two measures of decision errors (Correct Result and Correct Detection) yield similar numbers of total remaining errors ($r(47) = 0.952$, $p < 0.001$), none of the interactions for Correct Detection are statistically significant.

**Table 5**

*Summary of ANOVAs Addressing Hypothesis 1 According to Correct Result*

| Classification of Errors by | Passage Difficulty | Interaction | ANOVA |
|---|---|---|---|
| Correct Result | Easy | Dictation Ability x Alerting Threshold | $F(1, 45) = 2.86$, $p = .098$, $\eta^2 = .06$ |
| | | Editing Ability x Alerting Threshold | $F(1, 45) = 4.46$, $p = .040$*, $\eta^2 = .09$ |
| Correct Result | Difficult | Dictation Ability x Alerting Threshold | $F(1, 45) = 1.07$, $p = .308$, $\eta^2 = .023$ |
| | | Editing Ability x Alerting Threshold | $F(1, 45) = 0.389$, $p = .536$, $\eta^2 = .009$ |
| Correct Detection | Easy | Dictation Ability x Alerting Threshold | $F(1, 45) = .723$, $p = .40$, $\eta^2 = .016$ |
| | | Editing Ability x Alerting Threshold | $F(1, 45) = 2.59$, $p = .114$, $\eta^2 = .054$ |
| Correct Detection | Difficult | Dictation Ability x Alerting Threshold | $F(1, 45) = 2.24$, $p = .142$, $\eta^2 = .047$ |
| | | Editing Ability x Alerting Threshold | $F(1, 45) = 1.68$, $p = .202$, $\eta^2 = .036$ |

\* Significant results at the 5% significance level

Figure 4 gives a visual representation of the significant interaction found between alerting threshold and editing ability (second row in Table 5). On easy passages, low verbals performed better (fewer total remaining errors) with the specific alerting threshold than with the sensitive one, while high verbals performed better using the sensitive alerting threshold. For comparison, Figure 4 also shows the equivalent results for difficult passages (fifth row in Table 5). We note that for difficult passages low verbals instead performed better with the sensitive alerting threshold (corroborating later results addressing Hypothesis 2), while high verbals were almost unaffected by threshold.

**Figure 4** *Average number of total remaining errors after editing*

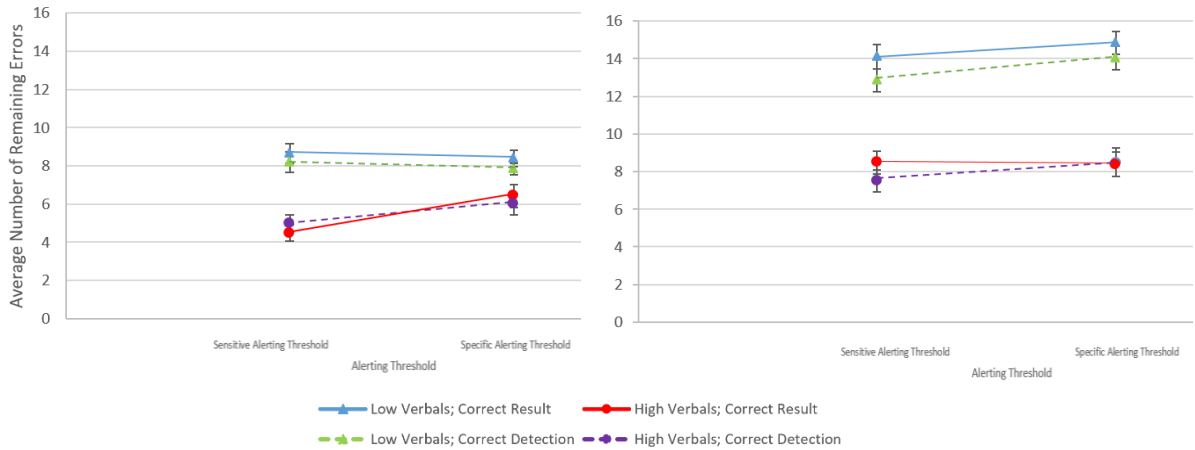

*Note.* Each line represents the performance of a certain user ability group (according to editing ability) when using the specific versus the sensitive alerting threshold (low verbals are represented by triangular markers and high verbals by round markers). Dashed lines represent performance based on Correct Detection while solid lines represent performance based on Correct Result.

**Results Addressing Hypothesis 2: Interaction of ability, threshold and difficulty**

To test H2, that user performance differs significantly between alerting thresholds depending on *both* user ability *and* passage difficulty (as indeed suggested by Figure 4), we conducted 2 x 2 repeated measures ANOVAs exploring the total remaining errors, including the within-subject variables alerting threshold (sensitive, specific threshold) and passage difficulty (easy, difficult). Ability was again included as a covariate. The significant three-way interactions (see Table 6) provide strong evidence for H2.

**Table 6**
*Summary of ANOVAs Addressing Hypothesis 2*

| Classification of Errors by | Interaction | ANOVA |
|---|---|---|
| Correct Result | Dictation Ability x Alerting Threshold x Passage Difficulty | $F(1, 45) = 4.62$, $p = .037$*, $\eta^2 = .093$ |
| | Editing Ability x Alerting Threshold x Passage Difficulty | |
| | | $F(1, 45) = 3.97$, $p = .052$, $\eta^2 = .081$ |
| Correct Detection | Dictation Ability x Alerting Threshold x Passage Difficulty | $F(1, 45) = 4.59$, $p = .038$*, $\eta^2 = .092$ |
| | Editing Ability x Alerting Threshold x Passage Difficulty | |
| | | $F(1, 45) = 6.64$, $p = .013$*, $\eta^2 = .129$ |

\* Significant results at the 5% significance level

**User Preference**

The final questionnaire asked participants which "CAT" they preferred; thirty-five expressed a preference. Results show that they were neither significantly more likely to choose the alerting threshold that was "better" nor the one that was "worse" (for their own performance) $x^2(1) = 1.46$, $p = 0.23$.

We also investigated whether either higher-ability users, or those to whom the "better" threshold gave greater advantage, were better at this choice. Figure 5 reveals no such patterns ($r(35) = -0.159$, $p < 0.363$): neither higher ability (according to dictation score) nor larger differences in the number of errors between the two thresholds (absolute y-values) resulted in more users "choosing well" (a higher number of positive y-values).

**Figure 5** *Users' ability to identify the alerting threshold with which they performed better*

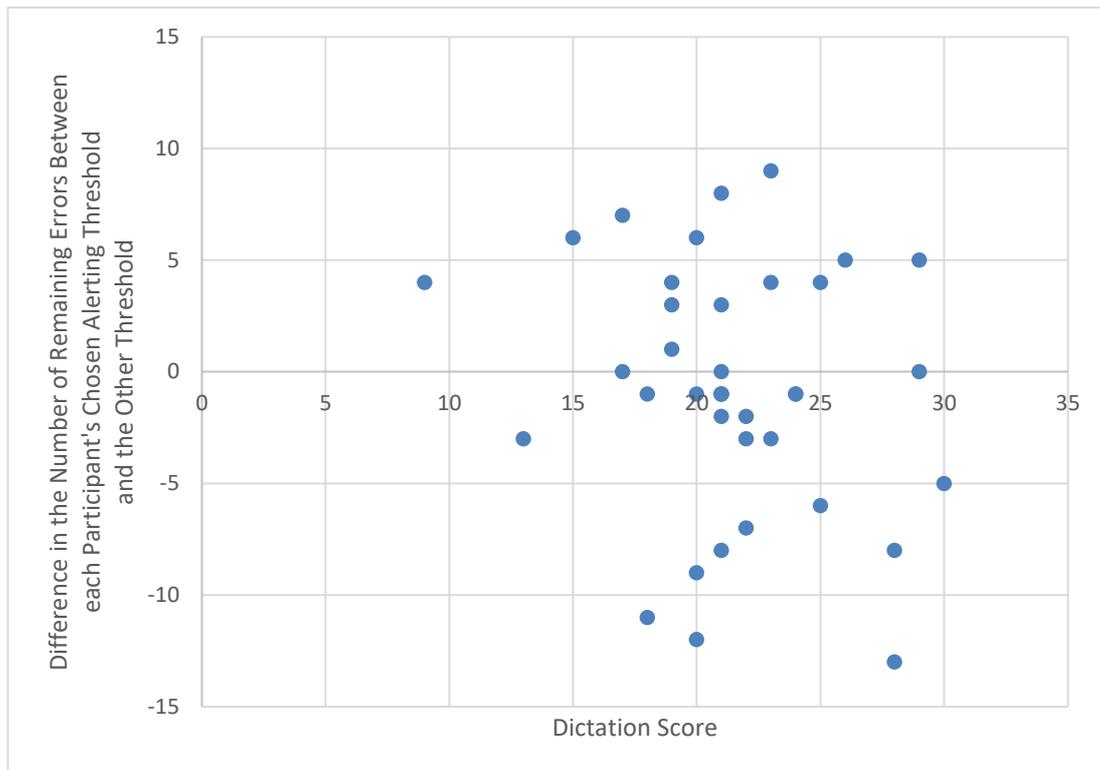

*Note:* Points above the y=0 axis indicate users who chose correctly

## Discussion

The results support both our hypotheses, H1: which alerting threshold leads to fewer decision errors depends on user ability; and H2: the "better" threshold also depends on the interaction between user ability and task difficulty. Our first hypothesis, that there would be an interaction between user ability and alerting threshold, received more limited support, only manifesting for the easy, not the difficult, text – and only for one of our two ability measures.

Why this difference between easy and difficult passages? One possible explanation can be derived from the work by Meyer et al., 2002, discussing "Why better operators receive worse warnings". Possibly, in "difficult" passages some words were difficult enough that even the "high verbals" could not decide correctly what to do, no matter whether or not CAT prompted them correctly: the hypothesized difference (advantage of the more sensitive threshold for the high verbals) was not present. For both user categories, more correct prompts failed to help, while more false positives harmed performance. Thus, changing the threshold had no detectable effect (Figure 4) with difficult passages. This observation supports our thesis: how best to adjust thresholds should be determined by empirical measurements involving each specific CAT on realistic samples of its users and decision tasks.

We find not only factors that separately affect which alerting threshold is most appropriate, but more interestingly, interactions among them. An alerting threshold that suits one user may not be best for another, and the same user may perform better with different alerting thresholds depending on task difficulty. To our knowledge, this is the first experimental confirmation of these effects.

The effects we observed on user errors from changing the alerting threshold were non-negligible. In one condition (Figure 4, "high verbals (editing ability)"; "easy passage"; "Correct Result"), the number of user errors differed between the two alerting thresholds by 30% ($t(21)$= -3.52, $p$ = 0.002) – with the same numbers of misspellings and CAT errors. The CAT thresholds had not been specifically tailored for these users and tasks; so, intentional tailoring could reduce errors even more. If such effects occurred in more critical applications, where errors cost lives and/or other harm, they would be of great interest. These results establish that suboptimal alerting can make users – even high-ability users – commit significantly more decision errors.

These results can be further interpreted as evidence that adjustable alerting thresholds can mitigate "automation bias", or as we prefer to refer to the phenomenon: "automation mediated errors". We hope this work will encourage other researchers in the area of automation bias to recognize adjustable thresholds as a potential solution to some automated-mediated errors, especially since reviews in this area (even recent ones) (Goddard et al., 2011; Lyell & Coiera, 2016) do not mention it. We also note that even in the area of adaptable/adaptive automation where the issue has been discussed (Kidwell et al., 2012), the focus is often on varying the functions delegated to automation ("LOAs") rather than varying alerting thresholds.

As the results show a significant improvement in user decisions, the natural question is then: how should the appropriate threshold be chosen for each user/task? This is an important question because many devices already marketed with adjustable thresholds are not released with guidelines for adjusting them. Results from our experiment support other researchers' findings that letting users choose does not always guarantee correct choices. Users in our study did not appear able to identify which alerting threshold helped them perform better. Meyer & Sheridan (2017) take this further by raising a question regarding "the effects of expertise on the ability to adjust thresholds." We found no such effects. Perhaps in some CAT applications users could be given frequent feedback about their performance to improve their choices of alerting threshold. However, this has not been conclusively reported to be effective. We suggest that suppliers of CATs with adjustable thresholds, and organizations using them, should instead explore other methods for selecting appropriate thresholds for each user/task, especially methods that rely on experimenting with a specific CAT on potential users and representative tasks to reveal empirical results that can be extrapolated to provide appropriate threshold guidelines.

A more advanced method would be designing adaptive CATs that adjust their own thresholds. For example, Meyer & Sheridan (2017) suggest that a CAT for neonatal intensive care could automatically adjust its threshold as the baby gains weight. Automatic adjustment to the difficulty of the decision, and to the user's ability, are intriguing, though challenging directions for innovation.

Apart from the direct results highlighted above, we also discussed methodological improvements to this category of studies by highlighting how different measures of ability (editing vs. dictation) and decision accuracy (Correct Detection vs. Correct Result), can change results. We also discussed how such distinctions can have practical implications in critical CAT applications by referring to the example of cancer screening in mammography (Alberdi et al., 2008).

Limitations of our study include that we studied one, non-critical use of CATs, spell-checking, and in an artificial setting. However, the results justify exploring the potential of adjustable thresholds in more critical CAT applications. Experience, as mentioned earlier, suggests that similar cognitive mechanisms with similar effects may operate there. Another limitation concerns our measure of users' abilities to correctly choose the threshold that is "better" for their performance. In our experiment, participants did not actually choose the thresholds, but only stated their preferred threshold after the fact, and using only their observations of the CAT's behavior. It is worth exploring whether feedback about their (and the CAT's) performance, or other relevant information, could improve users' choices.

Despite limitations, our findings imply strong suggestions for practice: adjusting alerting thresholds may seriously improve performance of CAT users, but it may only achieve this benefit if coupled with effective guidelines for choosing the appropriate threshold, especially as the choice cannot always be left to the users. The extent to which adjustable thresholds reduce errors will vary between CAT applications, but their potential seems such that it should be investigated, where affordable. We recommend that CAT developers and user organizations experiment in their specific application domains about: (1) whether adjustable thresholds offer the potential substantial improvements there and (2) how the threshold can be chosen in practice, depending on the task, and on the users. Even developers of devices with a single threshold may benefit from this work by considering, when choosing this threshold, how it will depend on the expected users' abilities and the ranges of task difficulties. Safety regulators and safety standard writers should also consider guidelines to inform the selection of appropriate alerting thresholds, essential for effective application.

Further research exploring alerting threshold as a function of other factors beyond user ability and task difficulty is also needed, as well as determining how many alerting thresholds to use, the set of factors driving the choice, and how the change between thresholds should happen. With many CATs on the market already featuring adjustable alerting thresholds, and more CAT developers considering their introduction, more work in this area will be needed.

**Acknowledgments**: The authors would like to thank Dr. Eugenio Alberdi and Dr. Andrey Povyakalo for their insightful advice, helpful discussions, and valuable input to this research.